\documentclass[twocolumn,aps,pre,showpacs,preprintnumbers]{revtex4}
\RequirePackage{ifpdf} 
\usepackage[dvipdf]{graphicx}
\usepackage{epstopdf}
\usepackage{epsfig} 
\usepackage{hyperref}
\usepackage{graphics}
\usepackage{color}      
\usepackage{dcolumn}

\begin{document}
\title{The physics of  Erythrocyte Sedimentation  Rate }

\author{Mesfin Asfaw Taye}
\affiliation{West Los Angles College, Science Division \\9000  Overland Ave, Culver City, CA 90230,USA }

\begin{abstract} 

An erythrocytes  sedimentation rate (ESR) measures  how fast a  blood sample sediments along a test tube in one hour  in a clinical laboratory. Since elevated level of  ESR is associated  with inflammatory diseases,  ESR is one of the routine hematology test   in  a clinical laboratory. In this paper,  
the physics of erythrocyte (RBC)   sedimentation  rate  as well as the dynamics of the RBC is explored by modeling  the dynamics of the cells as the motion of Brownian  particle moving in a viscous  medium. The viscous friction of blood $\gamma$ is considered  to decrease as the temperature of the medium increases \cite{aa1}. The results obtained in this work show that 
the ESR increases as the number of red blood cells (that bind together in the sedimentation process) steps up.  The room temperature also affects  the sedimentation rate. As the room temperature rises up, the ESR steps up.  Furthermore the dynamics of the RBC  along  a Westergren pipet  that is held in  an upright position is explored.  The exact analytic result depicts that  the velocity of cells  increases  as  the number of cells  that form  rouleaux steps up. Since   our study is performed by considering real physical parameters, the results obtained in this work non only agree with the experimental observations but also helps to understand most hematological experiments that are conducted in vitro.    
\end{abstract}
\pacs{Valid PACS appear here}
\maketitle
 
\section {Introduction}

An erythrocyte (red blood cell)  sedimentation rate  measures  how fast a  blood sample sediments along a test tube in one hour in a clinical laboratory as shown in Fig. 1.  This common hematology test is performed by  mixing the whole blood with anticoagulant \cite{aa2,aa3,aa4}.    The blood is then placed in an upright  Wintrobe or Westergren tube. The sedimentation rate of  the red blood cells (ESR)  is measured in millimeters (mm) at the end of one hour.  The  normal ESR   varies from 0-3mm/hr for men and 0-7 mm/hr for women \cite{aa2,aa3,aa4}.  High ESR   is  associated with diseases that cause  inflammation. Thus the ESR lab test helps to diagnose certain medical problems. 
\begin{figure} 
 \includegraphics[width=6cm]{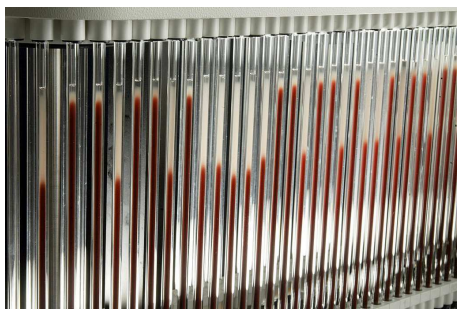}
\caption{  Figure that shows  the sedimentation  rate  of  RBC  on Westergren pipet \cite{aa5}. 
This  hematology test is performed  by  mixing  whole blood with anticoagulant.  The blood is then placed in an upright  Wintrobe or Westergren tube. The  sedimentation rate of the red blood cells is measured in millimeters (mm) at the end of one hour. }
\end{figure}

One can ask why   the ESR   is higher  in case of inflammatory  diseases. This question can be answered by considering all the forces acting  on the red blood cells (RBCs).    Consider a blood sample that is  placed in an upright  Wintrobe or Westergren tube.  The red blood cells in the blood sample are negatively charged and they tend to  repel each other.  Since the mass of a single RBC is too small, its gravitational force is too small to overcome the viscous  friction force of the blood.  As a result,  the RBCs  remain   in the blood sample without being precipitated.  In case of  an inflammatory disease, the blood level of  fibrinogen becomes too high \cite{aa4, aa6}. The presence of  fibrinogen  forces the RBCs to stick  each other  and as a result they  form  aggregates of RBC called rouleaux as depicted in Fig. 2. As the mass of the  rouleaux increases, the weight of the  rouleaux  dominates the vicious friction and as a result, the RBCs  start to precipitate. 
\begin{figure} 
 \includegraphics[width=6cm]{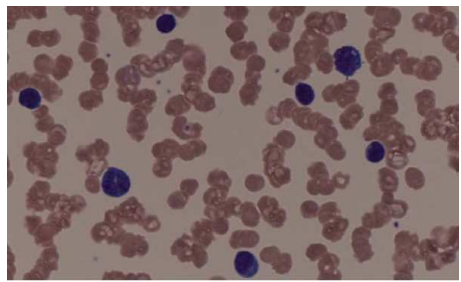}
\caption{ Rouleaux formation of RBC  \cite{aaa5}.}
\end{figure}

Elevated erythrocyte sedimentation rate is not only due to high level of  fibrinogen  and gamma globulins but also can be caused by  inclination of the test tube.  The temperature of the laboratory (blood sample)  also significantly affects the test result. As the temperature  of the sample  steps up, the ESR increases.   False negative results are also  observed  due the  abnormally shaped RBC, low room  temperature and  use of excessive  anticoagulant  during blood sampling.

In this paper we study the ESR and the dynamic of red blood cells analytically. Since RBC is microscopic in size, its dynamics can be model as a Brownian particle walking in a viscous medium.  As  blood is a highly viscous medium, the chance for the RBC to accelerate is negligible. One can then  neglect the inertia effect  and the corresponding dynamics can be studied via Langevin equation or Fokker Planck equation \cite{aa7,aa8,aa9,aa10,aa11}. Solving the Fokker Planck equation analytically, we explore how  the sedimentation rate behaves as a function of the model parameters. The  exact analytic  results indicate that the ESR increases as the red blood cells   form aggregates. Particularly the result obtained in this work exposes  the rate of sedimentation as a function of the number of red blood cells that bind together in the sedimentation process. Since   our study is performed by considering real physical parameters, the results obtained in this work non only agree with the experimental observations but also helps to understand most hematological experiments that are conducted in vitro.   

The room temperature  also  considerably affects the outcome of the ESR.  
Various
experimental studies showed that the viscosity of the
fluids tends to decrease as the temperature of the
medium increases \cite{aa1, aa8}. This is because increasing the temperature
steps up the speed of the molecules and this
in turn creates a reduction in the interaction time between
neighboring molecules. As a result, the intermolecular
force between the molecules decreases and hence the
magnitude of the viscous friction decreases. Our analysis indicates that as the room temperature  steps  up, the ESR increases supporting the previously observed experimental analysis. Not only  did we reconfirm the previously known  results, but also  propose a way of controlling false positive or false negative results.

Furthermore we study the dynamics of the RBC  along  a Westergren pipet  that is held in an upright position.  The exact analytic result depicts that  the velocity of the RBC  increases  in time. As the number of cells  that form  rouleaux steps up,  the velocity increases since the weight of the cluster dominates the viscous friction force.  Moreover,  when the temperature of the room increases, the velocity steps up since  the viscosity  of the fluid tends to decreases with temperature.  On the other hand, the position of the cells along the tube is investigated as a function of time and cluster size.

The rest of the paper is organized as follows. In section II, we present the model system. In section III, we explore the dependence of ESR on the number of RBCs that form clusters as well as on the background temperature. The dynamics of the cells is explored in section IV. Section V deals with summary and conclusion.

\section{The model}  

The dynamics of RBC is modeled as a Brownian particle that undergoes a biased random walk  on one dimensional upright Westergren pipet length $L=200 mm$ under the influence  of gravitational  force 
\begin{equation}
  f=Nmg .
\end{equation}
where $m=27X10^{-12}$ is the mass of the red blood cells  and $g=9.8m/s^{2}$ is the gravitational acceleration. $N$ denotes the number of red blood cells that forms rouleaux. The total number of RBC in  Westergren pipet  can be inferred since in Westergren lab analysis, $2000 mm^3$ blood is collected  into the test tube that contains  $500 mm^3$ preservative called sodium citrate.  The normal value of RBC  on average  varies from $5X10^6-6X10^6/mm^3$ \cite{aa12,aa13,aa14}. This implies that the total number of the RBCs contained in Westergren tube varies as $N=10X10^9-12X10^9$.

{\it Overdamped case:\textemdash}
Since blood is a highly viscous medium, the chance for the RBC to accelerate is negligible. One can then  neglect the inertia effect  and the corresponding dynamics can be studied via Langevin equation 
\begin{eqnarray}
\gamma{dx\over dt}&=& -f + \sqrt{2k_{B}\gamma T}\xi(t).
\end{eqnarray}
For  a non-Newtonian  fluid such blood, it is reasonable  to assume  that   
when  the  temperature of the blood sample  increases by $1$ degree celsius, its viscosity  steps down by $2$ percent \cite{aa15} as 
 $\gamma  =    B-{2B\over 100}(T-T^R)$
where $B=4 kg/s$ is the viscosity of blood at a room temperature ($T^R=20$ degree celsius) and $T$ is the temperature \cite{aa16}.  
 The random noise $\xi(t)$ is assumed to be Gaussian white noise 
satisfying the relations $\left\langle  \xi(t) \right\rangle =0$ and $\left\langle \xi(t)  \xi(t')\right\rangle=\delta(t-t')$. $k_{B}=1.38064852 × 10^{-23} m^{2} kg s^{-2} K^{-1}$ is  Boltzmann constant.

In the high friction limit, the dynamics of the Brownian particle is governed by 
\begin{equation}
{\partial P(x,t)\over \partial t}={\partial\over  \partial x}
\left[{f\over \gamma}P(x,t)+{\partial \over \partial x}\left({k_{B}T\over \gamma}P(x,t)\right)\right]
\end{equation}
where $P(x,t)$ is the probability density of finding the particle (the cell) at position $x$ and  time $t$.

The RBC  hops in a periodic  isothermal medium  of length $L=200mm$. The cell is also exposed to the  external load. In order to calculate the desired thermodynamic quantity, let us first find the probability distribution. After some algebra one finds the   probability distribution  as 
\begin{eqnarray}
P(x,t)&=&\sum_{n=0}^\infty \cos[{n\pi \over L}(x+t {f\over \gamma})]e^{-({n\pi  \over L})^2t{k_{B}T\over \gamma}}
\end{eqnarray}
where  $T$ is the temperature of the medium. For detailed mathematical analysis, please refer to my previous work  \cite{aa16}.
The particle current is then given by 
\begin{eqnarray}
J(x,t)&=&-\left[f P(x,t) + k_{B}T{\partial P(x,t) \over \partial x}\right].
\end{eqnarray}
The velocity of the cells at any time is given by 
\begin{eqnarray}
V(x,t)&=&\int_{0}^{x}J(x',t)dx'
\end{eqnarray}
while the position of the cells can be found  via
\begin{eqnarray}
y(x,t)&=&\int_{0}^{x}P(x',t)x'dx'.
\end{eqnarray}

{\it Underdamped case :\textemdash} If  the inertia effect is included, the Langevin equation can be written as 
\begin{eqnarray}
{dV\over dt}&=& -\gamma V-f + \sqrt{2k_{B}\gamma T}\xi(t).
\end{eqnarray}
After some algebra, the average velocity is simplified to 
\begin{eqnarray}
V(t)=\left({1-e^{-{\gamma t\over m}}\over \gamma}\right)f.
\end{eqnarray}
At steady state (in long time limit), the velocity (Eqs. (6) and (9) ) approach $V=f/\gamma$. This result agrees with our previous  works \cite{aa8,aa17}.

The diffusion constant for the model system is given by  $D={k_BT\over \gamma}$. This equation   is valid when  viscous friction is  temperature dependent showing that the effect of temperature on the cells mobility 
is significant. When temperature increases, the viscous friction gets attenuated and as a 
result the diffusibility of the particle increases. Various experimental studies also showed
that the viscosity  of the  medium  tends to decrease as the  temperature of the medium 
increases \cite{aa1}. This is because increasing the temperature steps up the speed of the 
molecules, and this in turn creates   a reduction in the interaction time between
neighboring  molecules.  As a result, the intermolecular force between the molecules 
decreases and hence  the magnitude of the viscous friction decreases.
Next we explore the dependence of the ESR on  number of RBCs that form  rouleaux.

\section{The erythrocyte sedimentation rate in Westergren pipet  }

\subsection{ESR as  a function of the number of red blood cells that form rouleaux }

In this section, the dependence of ESR on the number of  erythrocytes that bind together in the sedimentation process is explored.  As discussed before,  the repulsive coulomb force between the RBCs  keeps apart the cells from binding. As the result the viscous friction of the  blood   compels the cells to remain   suspended  in the solution. The presence of inflammatory disease results in elevated plasma  fibrinogen.  The fibrinogen  forces the RBC to stick to each other to form  aggregates of RBC called rouleaux. As the mass of the  rouleaux increases, the weight of the  cluster  dominates the vicious friction and  as a result the RBC  starts to sediment.   As depicted  in the work \cite{aa18}, the sedimentation rate is linearly correlated with fibrinogen 
blood level.

As the fibrinogen blood level steps up, more RBCs tend to bind. Hence it is vital to  explore the number of  red blood cells that are involved in the sedimentation process at a given ESR. 
Exploiting Eqs. (6), (7) and (9), one can see that as the number of red blood cells ($N$) that form rouleaux steps up, the sedimentation rate increases (see Fig. 3).   Figure 3 depicts the plot of ESR as a function of the number of RBCs (at $20$ degree celsius)  that forms rouleaux.  As shown in the figure, the ESR increases  as $N$  steps up.  The abnormal value for ESR is  observed when $5X10^{5}$ or more  RBCs  form aggregate.
\begin{figure}[ht]
\centering
{
    \includegraphics[width=6cm]{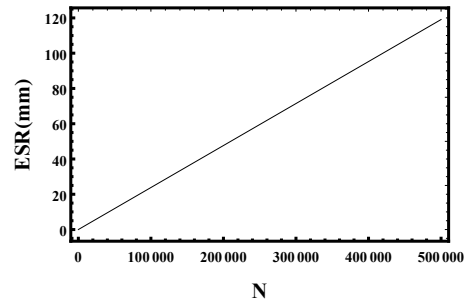}
}
\caption{ (Color online) Erythrocyte sedimentation  rate in one hour at $20$ degree celsius. The ESR steps up as the number of red blood cells ($N$)that form rouleaux  increases. The abnormal value for ESR is  observed when $5X10^{5}$ or more  RBCs  form aggregate.} 
\label{fig:sub} 
\end{figure}

\begin{figure}[ht]
\centering
{
    \includegraphics[width=6cm]{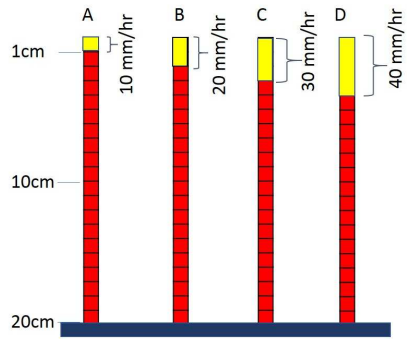}
}
\caption{ (Color online)  Schematic diagram that plotted based on the result shown in Fig. 3.  The figure shows that $ESR=10mm/hr$, $20mm/hr$, $30mm/hr$ and $40mm/hr$ when (the number of RBCs that form aggregate)  $N=4.1992X10^4$, $8.3984X10^4$, $N=12.5976X10^4$  and  $N=16.7968X10^4$, respectively.} 
\label{fig:sub} 
\end{figure}

Figure 3 can be more illustrated by drawing a schematic diagram  to show the dependence of $ESR$ on $N$. Based on the result  depicted  in Figure 3, we  redraw a schematic diagram in Fig. 4.  The figure shows that $ESR=10mm/hr$, $20mm/hr$, $30mm/hr$ and $40mm/hr$ when (the number of RBCs that form aggregate)  $N=4.1992X10^4$, $8.3984X10^4$, $N=12.5976X10^4$  and  $N=16.7968X10^4$, respectively. One can note that  each RBC  undergoes a biased random walk since the  external load (the weight of RBCs) compels the cells to sediment. The repulsive coulomb force interaction between the cells is negligible in comparison with the  gravitational force. Next  we will explore the effect of temperature on ESR. 

\subsection{The effect of temperature on ESR }

  The viscous friction  of the fluid  depends on the intensity of the background    temperature  of the fluid  showing that the effect of temperature on the cell mobility 
is significant. When temperature increases, the viscous friction gets attenuated and as a 
result,  the diffusibility of the cells increases as shown in the works \cite{aa1, aa8}. 
 For  a non-Newtonian  fluid such blood, it is reasonable  to assume  that   
when  the  temperature of the blood sample  increases by $1$ degree celsius, its viscosity  steps down by $2$ percent.  Exploiting Eqs. (6),   (7) and (9), the dependence of the  erythrocyte sedimentation  rate  as a function of temperature (in degree celsius) is depicted in Figure 5. In the figure, the number of RBCs that form clusters are fixed as $N=15000$, $N=10000$ and $N=5000$ from  top to bottom, respectively. The figure depicts that  the ESR steps up as the number of red blood cells ($N$) that form rouleaux  increases as well as when the temperature of the room steps up.
The same figure  shows that  up to 3mm/hr sedimentation rate difference  can be observed when the temperature of the room varies from 20 to 45 degree celsius.

To explore  the role of  the model parameters in detail, it is instructive  to draw the phase  diagram $N$ as a function of $T$  by fixing the sedimentation  rate  as shown in Fig. 6. In the figure,  the sedimentation  rates are fixed as $ESR=15mm$, $10mm$ and  $5mm$  from top to bottom, respectively. The phase diagram  exhibits that the temperature of the medium significantly affects the sedimentation rate. 

All of the above  analysis points that  the increase in the room temperature (the temperature of the sample) results in a false positive result. The  shape of RBC, plasma viscosity and inclination of the test tube  also affect the magnitude of ESR. Low hematocrit level is observed in anemic patents. As the hematocrit level decreases, the viscosity of the blood decreases which results in high ESR level. Excessive use of anticoagulant reduces the viscosity of the blood and in this case a false positive result can be observed.    

\begin{figure}[ht]
\centering
{
    \includegraphics[width=6cm]{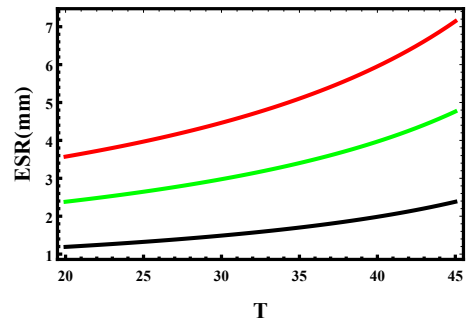}
}
\caption{ (Color online) Erythrocyte sedimentation  rate in one hour as a function of temperature in degree celsius. The number of RBCs that form clusters are fixed as $N=15000$, $N=10000$ and $N=5000$ from  top to bottom, respectively. 
The ESR steps up as the number of red blood cells ($N$)that form rouleaux  increases as well as when the temperature of the room steps up.} 
\label{fig:sub} 
\end{figure}

\begin{figure}[ht]
\centering
{
    \includegraphics[width=6cm]{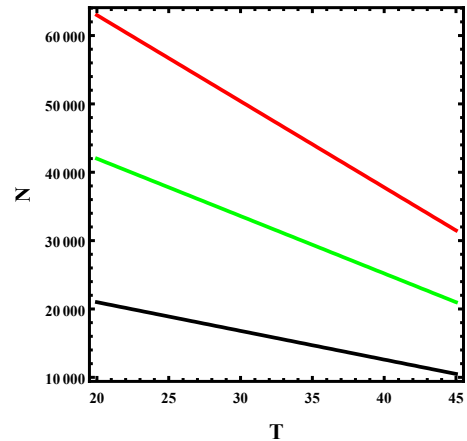}
}
\caption{ (Color online) The phase diagram $N$ as a function of $T$.   The sedimentation  rates are fixed as $ESR=15mm$, $10mm$ and  $5mm$  from top to bottom, respectively. The phase diagram  exhibits that the temperature of the medium significantly affects the sedimentation rate. 
} 
\label{fig:sub} 
\end{figure}

\section{The erythrocytes dynamics along Westergren pipet  }  

Once the blood is mixed with anticoagulant and allowed to stand in Westergren pipet, the red blood  cells start undergoing a biased Brownian motion against the viscous medium. Here care must be taken since the viscosity of the blood is also sensitive to the amount of anticoagulant administered  during the experiment. Excessive use of anticoagulant results in a false positive result. 

\begin{figure}[ht]
\centering
{
    \includegraphics[width=6cm]{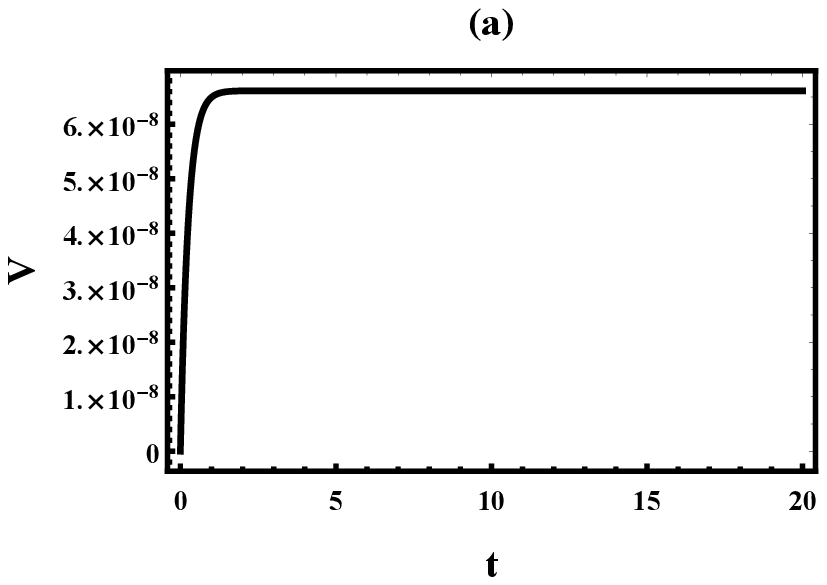}
}
\hspace{1cm}
{
    \includegraphics[width=6cm]{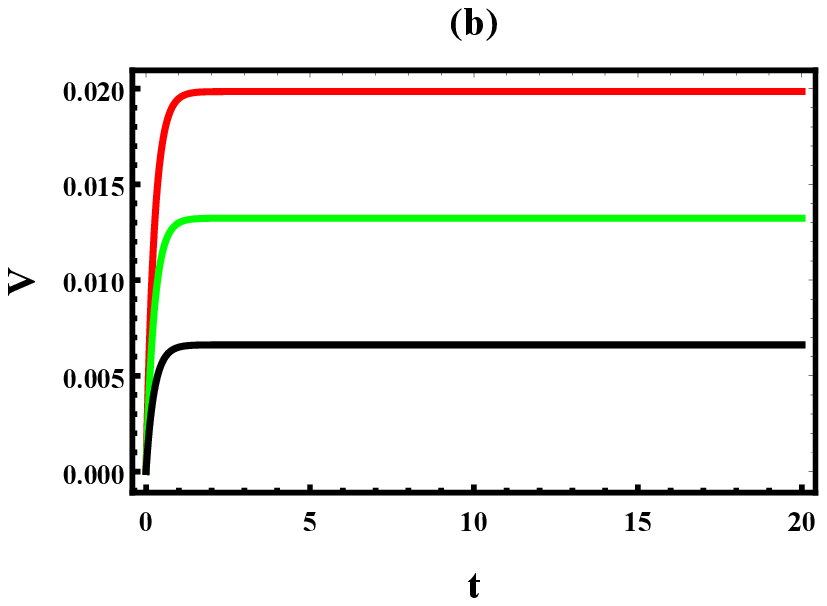}
}
\caption{ (Color online)(a) The velocity ($V (mm/s)$) of RBC as a function of time $t$ for a single RBC $N=1$ at $20$ degree celsius.
 (b)  The velocity ($V(mm/s)$) of RBC as a function of time $t$ for a cluster of RBCs $N=300000$, $N=200000$ and $N=100000$ from top to bottom at $20$ degree celsius, respectively.} 
\label{fig:sub} 
\end{figure}

As discussed before, the background temperature of the fluid compels the cells to walk randomly as long as the plasma fibrinogen level is above  a certain threshold. In this case,  the RBCs  repeal each other as they are negatively charged. When the blood fibrinogen level steps up, the cells form rouleaux as a result the gravitational force (the weight of the cluster) overcome the viscous friction. The cells then move with non-zero velocity as shown in Fig. 7. Figure 7a depicts  the velocity ($V$)  as a function of time $t$ for a single RBC $N=1$ at $20$ degree celsius. Figure 5b exhibits the velocity ($V$)  as a function of time $t$ for a cluster of RBCs $N=300000$, $N=200000$ and $N=100000$ from top to bottom at $20$ degree celsius, respectively. The figures exhibits that the  velocity steps up as $t$ and $N$ increase as expected. 
\begin{figure}[ht]
\centering
{
    \includegraphics[width=6cm]{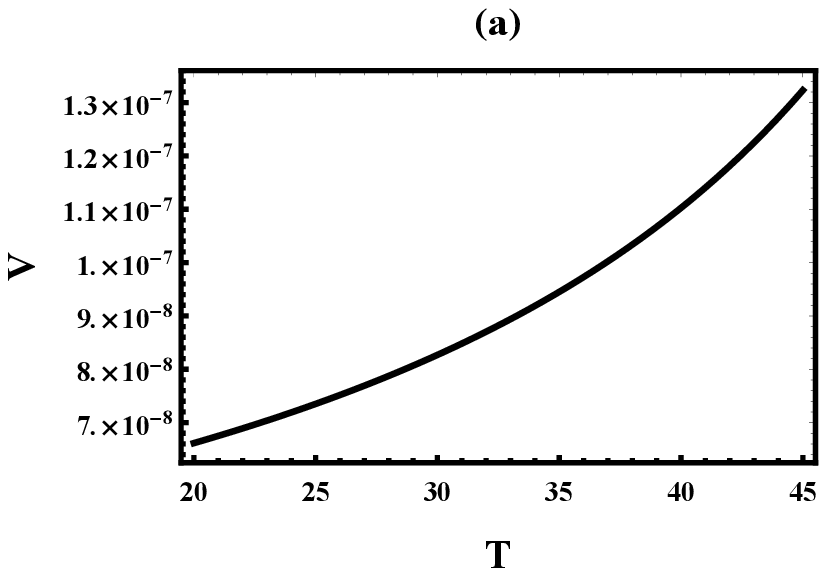}
}
\hspace{1cm}
{
    \includegraphics[width=6cm]{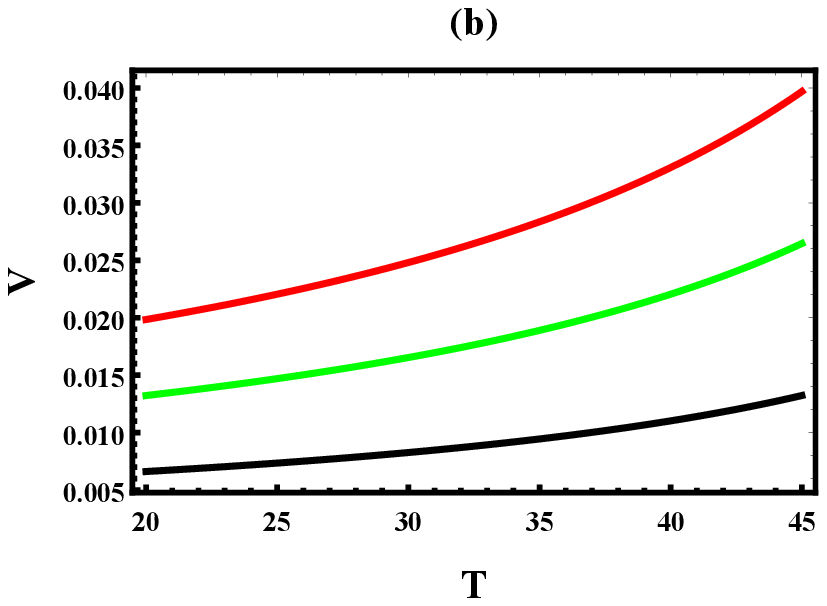}
}
\caption{ (Color online)(a) The velocity ($V(mm/s)$)   as a function of temperature (in degree celsius) $T$ for a single RBC $N=1$ at $t=600s$.
 (b)  The velocity ($V(mm/s)$) as a function of temperature (in degree celsius) $T$  for clusters of RBCs $N=300000$, $N=200000$ and $N=100000$ from top to bottom, respectively  at $20$ degree celsius.} 
\label{fig:sub} 
\end{figure}
Exploiting Eq. (6), one can see that at steady state (in long time limit), the velocity approach 
\begin{eqnarray}
V&=&f/\gamma. 
\end{eqnarray}
This result agrees with our previous works \cite{aa8,aa17}.

The intensity of the background temperature also significantly affects the velocity since  the  magnitude  of the viscous friction depends on the intensity of the temperature.   As the temperature steps up, the velocity increases (see Fig. 6) showing that this false positive result can be fixed by performing the clinical experiment at room temperature. Figure 8a shows the  velocity ($V$)  as a function of temperature (in degree celsius) $T$ for a single RBC $N=1$ at  fixed time $t=600s$. Figure 8b exhibits 
 the velocity ($V$)  as a function of temperature (in degree celsius) $T$  for clusters of RBCs $N=300000$, $N=200000$ and $N=100000$ from top to bottom, respectively. The temperature is fixed   at $20$ degree celsius. 
\begin{figure}[ht]
\centering
{
    \includegraphics[width=6cm]{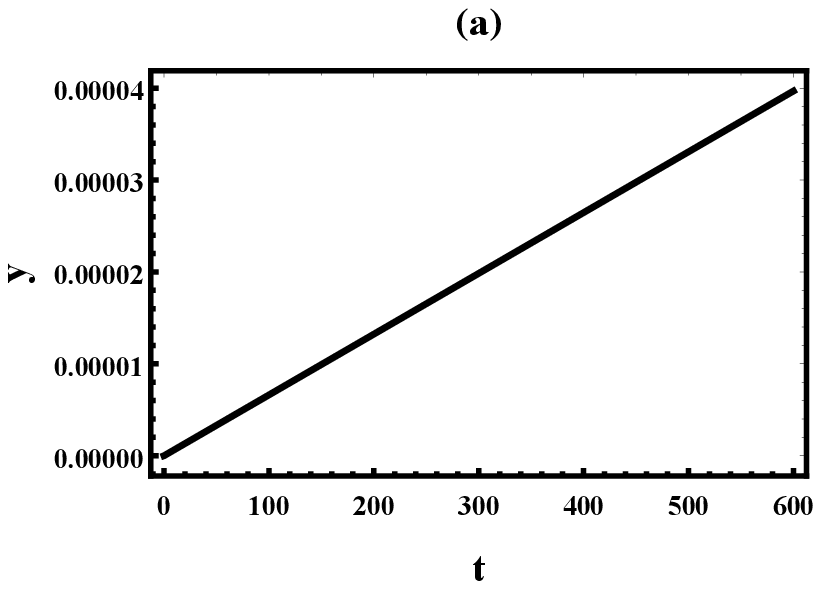}
}
\hspace{1cm}
{
    \includegraphics[width=6cm]{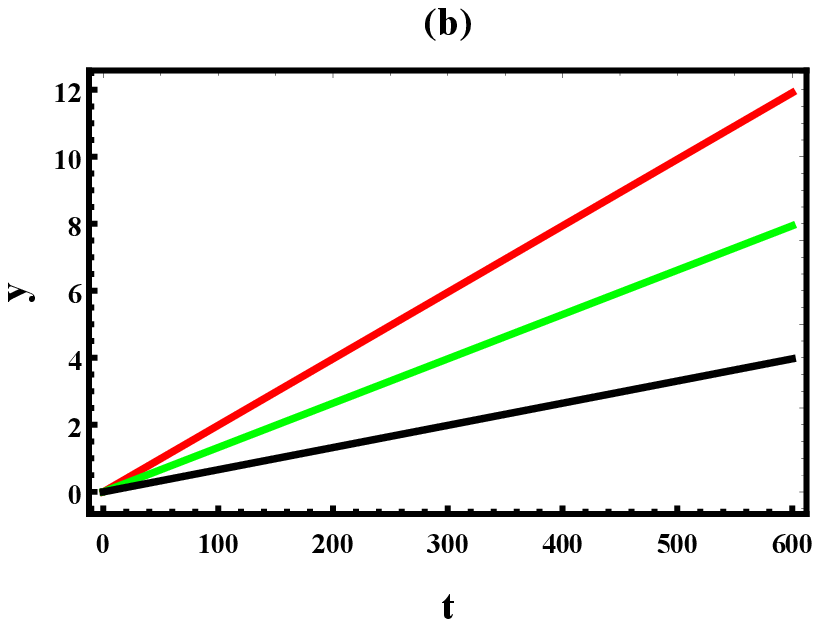}
}
\caption{ (Color online)(a) The displacement  ($x(mm)$) as a function of time $t$ for a single RBC $N=1$ at $20$ degree celsius.
 (b)  The displacement  ($x(mm)$) as a function of time $t$ for  clusters of RBCs $N=300000$, $N=200000$ and $N=100000$ from top to bottom, respectively  at $20$ degree celsius.} 
\label{fig:sub} 
\end{figure}
In Fig. 9a, the displacement  ($x$) as a function of time $t$ is  plotted  for a single RBC $N=1$ at $20$ degree celsius. Figure 9b depicts the displacement  ($x$) as a function of time $t$ for  clusters of RBCs $N=300000$, $N=200000$ and $N=100000$ from top to bottom, respectively  at $20$ degree celsius.

\section {Summary and conclusion}

An erythrocyte sedimentation rate is a common hematology test  which is performed by mixing the whole blood with anticoagulant.    The blood is then placed in an upright  Wintrobe or Westergren tube. The sedimentation rate of  the red blood cells  is measured in millimeters (mm) at the end of one hour.  ESR measures how fast a  blood sample sediments along a test tube in one hour in a clinical laboratory.  Elevated erythrocyte sedimentation rate is not only due to high level of  fibrinogen  and gamma globulins but also it can be caused by the inclination of the test tube.  The temperature of the laboratory (blood sample)  also significantly affects the test result.
 
In this work, we explore how  the ESR and the dynamic of red blood cells behave   analytically. Solving the Fokker Planck equation analytically, we explore the dependence of the ESR   as a function of the model parameters. The exact analytic  results indicate that the ESR increases as the red blood cells   form aggregates. Particularly the result obtained in this work exposes  the dependence of the rate of sedimentation as a function of the number of red blood cells that bind together in the sedimentation process. The effect of temperature on the sedimentation  rate is also studied  analytically. Our analysis indicates that as the  temperature pf  the room  steps  up, the ESR increases supporting the previously observed experimental analysis.
 
The dynamics of the RBC  along  a Westergren pipette  that held in upright position is also explored.  The exact analytic result depicts that  the velocity of the RBC  increases  in time. As the number of cells  that form  rouleaux steps up,  the velocity increases since the weight of the cluster dominates the viscous friction force.  Moreover, when the temperature of the room increases, the velocity steps up since  the viscosity  of the fluid tends to decreases with temperature.  On the other hand, the position of the cells along the tube is investigated as a function of time and cluster size.
 
In conclusion, in this work we present an important model system.  Since   this study  is performed by considering real physical parameters, the results obtained in this work non only agree with the experimental observations but also helps to understand most hematological experiments that are conducted in vitro.    Not only we reconfirmed the previously known  results, but also  we propose a way of controlling false positive or false negative results.

 {\it Acknowledgment.\textemdash} 
I would like to thank Mulu Zebene and Blyanesh Bezabih for the constant  encouragement.



\end{document}